\documentstyle[epsf,graphics,epsfig]{mn}

\newcommand{\asec}{^{\prime\prime}}

\title[Quasar Host Galaxies]{Two-Dimensional Modelling of Optical HST and 
Infrared Tip-Tilt Images of Quasar Host Galaxies}

\author[R.J. McLure et al.]
{R.J. McLure$^{1,2}$, J.S. Dunlop$^{1}$, M. J. Kukula$^{1}$
\\
$^{1}$Institute for Astronomy, University of Edinburgh, Blackford Hill, Edinburgh, EH9~3HJ\\
$^{2}$Nuclear and Astrophysics Laboratory, University of Oxford, Keble Road, Oxford, OX1 3RH}

\date{Accepted for publication in MNRAS}
\begin{document}
\maketitle

\begin{abstract}
A description is given of the method used to extract
quasar host-galaxy parameters from the deep Hubble Space Telescope ({\sc hst}) 
quasar images presented by McLure et al. (1999) and Dunlop et al. (2000). 
We then give 
the results of extensive testing of this technique on a wide range of
simulated quasar+host combinations spanning the redshift range of our
{\sc hst} study ($0.1 < z < 0.3$). These simulations demonstrate that,
when applied to our deep {\sc hst} images, our method of analysis can easily
distinguish the morphological
type of a given host galaxy, as well as determining its scalelength, 
luminosity, axial ratio and position angle to within an accuracy of a few 
percent. We also present new infrared tip-tilt images of 4 of the most luminous
quasars in our {\sc hst} sample, along with the outcome of modelling these data 
in a similar manner. The results provide further confidence in the 
accuracy of the derived host-galaxy scalelengths, and allow accurate
determination of $R-K$ colours for this subset of sources. All 4 of these 
quasar host galaxies have very similar red colours,
$R-K = 2.9 \pm 0.2$, indicative of a well-evolved stellar population. 

\end{abstract}


\section{Introduction}

In two companion papers (McLure et al. 1999; Dunlop et al. 2000) we present 
initial and final results from a deep Hubble Space Telescope ({\sc hst}) imaging 
survey of radio-quiet quasars (RQQs), radio-loud quasars (RLQs) and radio 
galaxies (RGs). The results presented in these papers were derived from the 
{\sc hst} images using a two-dimensional modelling technique developed to cope
with such complications as central image saturation, the undersampled nature 
of Wide Field (WF) camera images, accurate image centering, and 
the precise form of the {\sc hst} point spread function (PSF).
The primary purpose of this paper is to provide a description of 
this image analysis method, and to present the results of extensive testing on
simulated active-nucleus+host-galaxy images constructed to span the full range
of parameter space, and to mimic as closely as possible the real {\sc hst} data.
These tests on simulated data were central to the development of our modelling
technique, and also provide a means of estimating the typical errors in the
derived host galaxy parameters as a function of redshift.

We also present new data in the form of infrared tip-tilt images of 4 of the 
most luminous quasars in our {\sc hst}-imaging sample, and give the results of 
applying our two-dimensional modelling technique to these new $K$-band data. 
The results provide further confidence in our analysis technique, and allow us
to determine accurate $R-K$ colours for the hosts of these 4 quasars.

The layout of the paper is as follows. In section 2 we briefly review
the main modelling algorithm (for details see Taylor et al. 1996), and 
explain how the various 
problems which are specific to deep {\sc hst} imaging were tackled. In
section 3 we explain
how synthetic {\sc hst} quasar+host images were constructed to allow tests 
of the ability of the modelling technique to reclaim reliable host-galaxy 
parameters as a function of redshift and nuclear:host ratio. We then present
the results of these tests before proceeding in sections 4 and 5 to describe
extensions to the modelling algorithm which were introduced to avoid 
having to make the assumption that the host galaxies were either perfect
de-Vaucouleurs spheroids, or Freeman discs. The new UKIRT Tip-Tilt
IRCAM3 images of a subset of the quasar sample are presented in section
6 along with the results of modelling these data in a manner basically 
identical to that used to model the {\sc hst} data. We then briefly describe the
outcome of comparing the UKIRT- and {\sc hst}-derived galaxy parameters for these 
objects, before summarizing our main conclusions in section 7.
\section {Two-Dimensional Modelling of Quasar Host Galaxies}
\label{modeltech}
Given the difficulties associated with one-dimensional analysis
techniques it was decided that a fully two-dimensional approach was needed
to fully exploit the depth and resolution of the new {\sc hst} and UKIRT 
host-galaxy data. The two-dimensional modelling code used in the
analysis described in this paper, McLure et al. (1999) and
Dunlop et al. (2000), is a development of that which was originally designed 
for the analysis of the IRCAM 1 $K$--band imaging of this sample 
(Dunlop et al. 1993, Taylor et al. 1996). Consequently, only a brief
outline of the general algorithm is supplied here, with those interested in a
detailed description being referred to Taylor et al. (1996).

The model host galaxies are initially constructed on a two-dimensional array
with the surface-brightness of each pixel described by either a de
Vaucouleurs $r^{1/4}$ law (de Vaucouleurs \&
Capaccioli 1979):

\begin{equation}
\mu(r)=\mu_{o}\rm{exp}\left[-7.67\left(\left(\frac{r}{r_{e}}\right)^{1/4}-1\right)\right]
\end{equation}
\noindent
or a Freeman disc law (Freeman 1970):
\begin{equation}
\mu(r)=\mu_{o}\rm{exp}\left(-\frac{r}{r_{o}}\right)
\end{equation}
\noindent
The model galaxies are constructed at much higher resolution
than the data plate-scale in order to accurately simulate the
rapidly varying surface brightness at small radii (particularly
important with the $r^{1/4}$ profile). Following re-sampling of the
model host galaxies, an arbitrarily large amount
of flux is added to the central pixel to represent the unresolved
nuclear contribution. In order to
produce a simulated observation this `zero-seeing' model is then
convolved with a high signal-to-noise observation of the instrumental
point spread function (PSF). The goodness of fit between this
convolved model and the actual data is
then tested on a pixel-by-pixel basis via the $\chi^2$ statistic,
with the parameters determining the form of the model being
 iterated to find the minimum $\chi^{2}$ solution.  The model
parameters which are left free during this minimization process are:  

\begin{itemize}
\item{The luminosity of the nucleus}
\item{The central brightness of the host galaxy}
\item{The scalelength of the host galaxy}
\item{The position angle of the host galaxy}
\item{The axial ratio of the host galaxy}
\end{itemize}

\subsection{The WFPC2 PSF}
\label{psf}

\begin{figure}
\setlength{\epsfxsize}{0.4\textwidth}
\centerline{\epsfbox{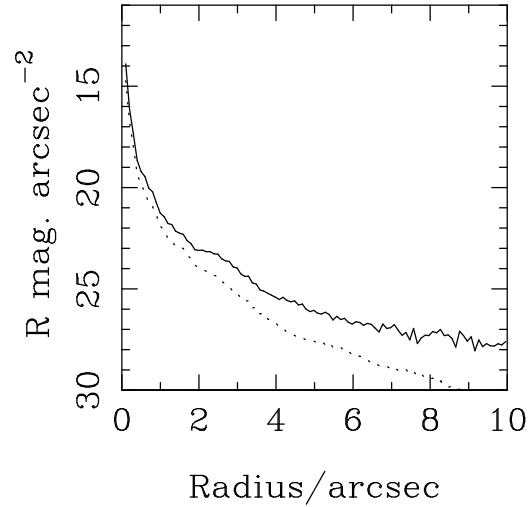}}
\caption{A comparison of our empirical F675W WF2 PSF (solid line) with
the equivalent {\sc tinytim} synthetic PSF (dashed line). Both have been
normalized to have the same central surface-brightness. It can clearly
be seen that the {\sc tinytim} model is unable to reproduce the halo
of scattered light outside a radius of $\simeq1.5\asec$.}
\label{psfcom}
\end{figure}

Many authors who have investigated host galaxies using the {\sc hst} (Hutchings
et al. 1994, Boyce et al.  1998, Hooper et al.
1997)  have made use of the
synthetic PSFs produced by the {\sc tinytim} software package (Krist
1998). This software is capable of reproducing the WFPC2 PSF
through any of the on-board filters, seemingly removing the need
to sacrifice valuable orbits observing the empirical PSF. However, the
inability of the {\sc tinytim} software to reproduce
the scattered light halo associated with the WFPC2 PSF (Krist 1998)
make these model PSFs unsuitable for the convolution of model
quasars. This fact is illustrated in Fig \ref{psfcom} which shows
a comparison between the high dynamic range empirical PSF used in our
analysis and the equivalent {\sc tinytim} model. It is clear from this
that if the model quasars do not include the scattered-light halo, the 
scalelengths of the underlying host galaxies could be seriously overestimated.

A further issue to be considered when analysing WFPC2 images is the 
severe undersampling of the WF CCDs. Given that the
central $1\asec$ of the $R$--band quasar images presented in McLure
et al. (1999) and Dunlop et al. (2000) are dominated by 
the unresolved nuclear component, it was imperative
that the undersampling issue was dealt with properly. In
order to deal with this problem a solution was arrived at
which, makes use of the sub-sampling capabilities of the {\sc tinytim}
software.  Although inadequate as a substitute for the
empirical PSF at large radii, the synthetic models produced by {\sc
tinytim} are an excellent match to the empirical PSF in the very central
regions where the optical effects are well understood.  The ability of
{\sc tinytim} to produce PSF models at fifty times higher resolution
than the WF2 plate-scale was used to
predict the appearance of the central nine pixels of the WF2 PSF with
any possible sub-pixel centring.  These
predictions are then $\chi^{2}$ matched to the core (central nine pixels) of
the quasar image to give centroiding with theoretical accuracy of $\pm
0.02$ pixels. Assuming that the {\sc
tinytim} model provides an accurate representation of the true PSF in
the central $\simeq0.5\asec$, it is then possible to
correct the sampling of the central regions
of the empirical PSF to the same sub-pixel centring as that of the
 quasar image.  
\subsection{Error Allocation}
\begin{figure}
\setlength{\epsfxsize}{0.3\textwidth}
\centerline{\epsfbox{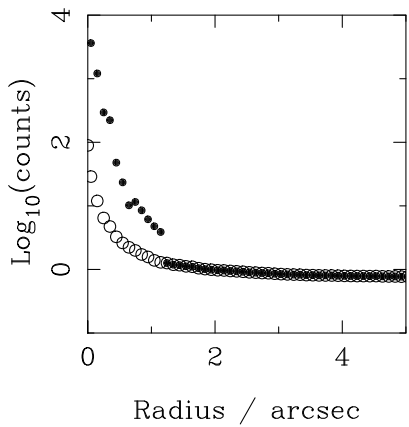}}
\caption{A typical {\sc {\sc hst}} error profile (0923+201). Shown in the 
figure are the predicted poisson errors (open circles) from the WF  
noise model, and the actual sampling errors
(filled circles), both calculated from azimuthal averaging in circular
annuli. It can be seen that outside a radius of $\sim 1\asec$ the
poisson and sampling errors are basically identical.}
\label{errprof}
\end{figure}
Regardless of the care with which model quasars are constructed,
two-dimensional modelling will fail to produce accurate
results if incorrect allocation of error weighting in the
$\chi^{2}$ test causes one area of the image
to dominate the fit at the expense of others. The minimum possible error that
can be associated with any pixel is a combination of the poisson error 
due to photon shot noise, plus the read-noise and dark current
contributions. Errors introduced during flat-fielding have proven to 
be negligible in both
the {\sc hst} data and the IRCAM 3 images considered in here. Figure
\ref{errprof} shows a comparison between poisson noise predicted by
the WF2 noise model (Biretta et al. 1996), and the actual 
statistical sampling errors as measured from a typical image. 

\begin{figure}
\setlength{\epsfxsize}{0.4\textwidth}
\centerline{\epsfbox{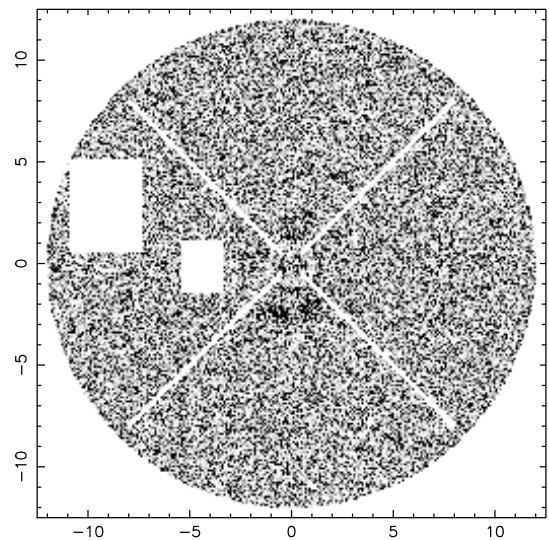}}
\caption{The reduced $\chi^{2}$ map for the best-fitting model to the
{\sc hst} image of the radio-loud quasar 1217+023. The grey-scale is a
linear stretch running
 between $0\rightarrow2$. It can be seen that a uniform spread of
 $\chi^{2}$ values has been achieved, with no area of the image
 dominating the fit. The blank
areas in the map are due to the masking from the fitting process of
two companion objects and the highly variable diffraction spikes.}
\label{chimap}
\end{figure}

As can be seen from Fig \ref{errprof}, the
errors calculated using the noise model are in near-perfect agreement
with what is actually seen from the data for all pixels outside a
radius of $\simeq 1\asec$ from the quasar core.  Consequently, all
pixels outside a radius of $1\asec$ are simply
allocated their expected poisson error during the model fitting
process. However, Fig \ref{errprof} also demonstrates that this system
would grossly underestimate the actual error introduced inside
$\simeq1\asec$ by the severe undersampling of the WFPC2 PSF.
Therefore, inside a radius of $1\asec$  a different system is used to
allocate the error weighting. A series of ten pixel-wide circular
annuli centred on the quasar are constructed, and the variance
($\sigma^{2}$) of the distribution of the pixels falling within each
annuli calculated. All of the pixels falling within a particular
annulus are then allocated the annulus variance as their error weighting.
 This procedure is still justified even when the underlying
host galaxy has a clear ellipticity since, even at {\sc hst} resolution, the
central $1\asec$ of the quasar images are dominated by the circular
symmetry of the PSF. Due to the success of the
re-sampling technique discussed above, it was possible to assign
a poisson error weighting to the central pixel without biasing the fitting
procedure. This complicated process has proven to be successful, with
the vast majority of the 33 {\sc hst} objects modelled having minimum
$\chi^{2}$ solutions lying in the $\nu\pm\sqrt{2\nu}$ region expected
for correct weighting (Dunlop et al. 2000). Confidence in the 
weighting system is further
strengthened by examination of the pixel maps of reduced $\chi^{2}$ values
produced for the best-fitting models. Shown in Fig \ref{chimap} a
typical $\chi^{2}$ map from the best-fit to the radio-loud quasar
1217+023. It is clear from this figure that the desired uniform spread of
$\chi^{2}$ values has been achieved, with no one area of the
image dominating the fit.   

\subsection{Minimization}
\label{minimum}
\begin{figure}
\setlength{\epsfxsize}{0.2\textwidth}
\centerline{\epsfbox{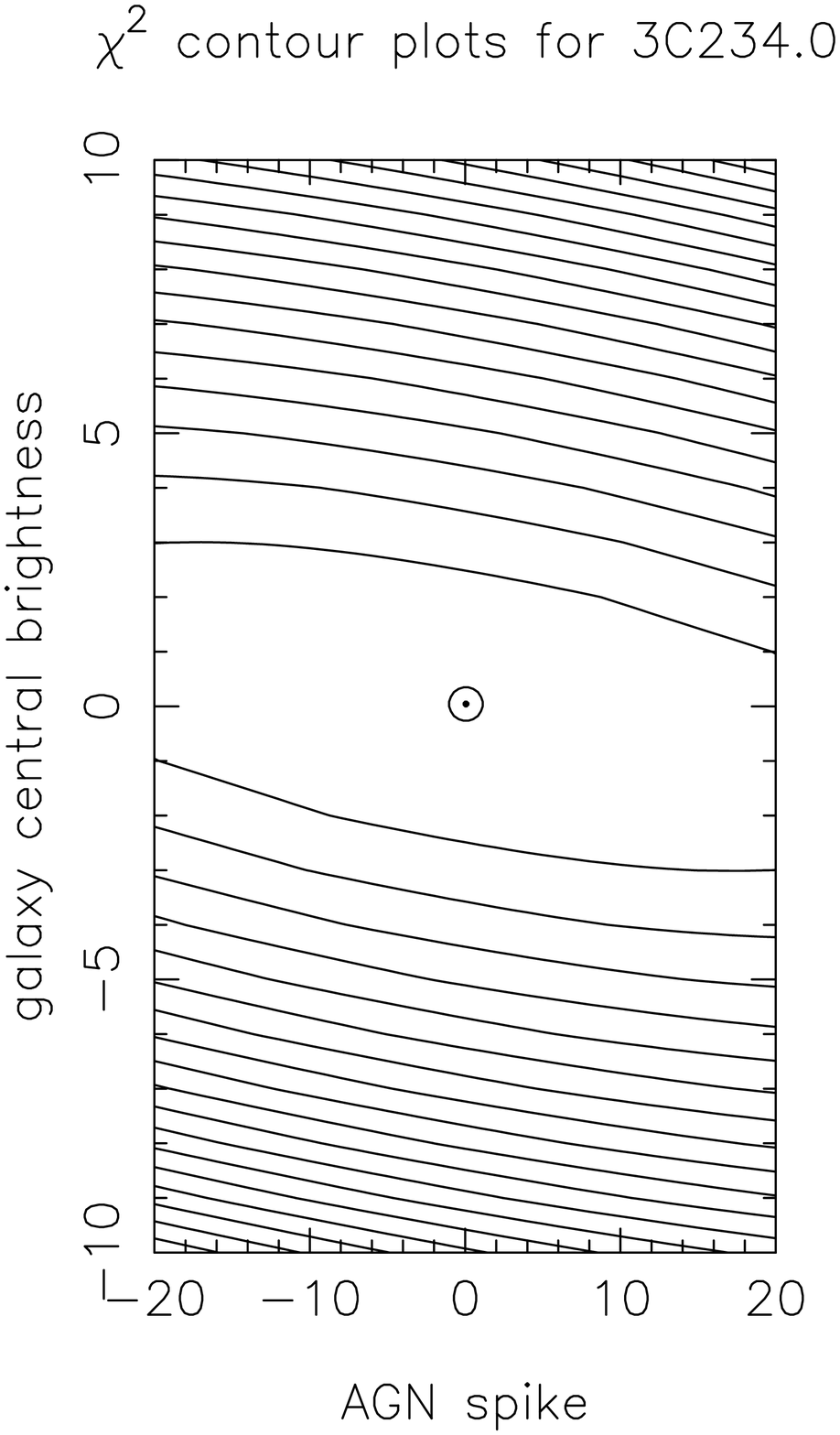}}
\vspace{0.25in}
\setlength{\epsfxsize}{0.2\textwidth}
\centerline{\epsfbox{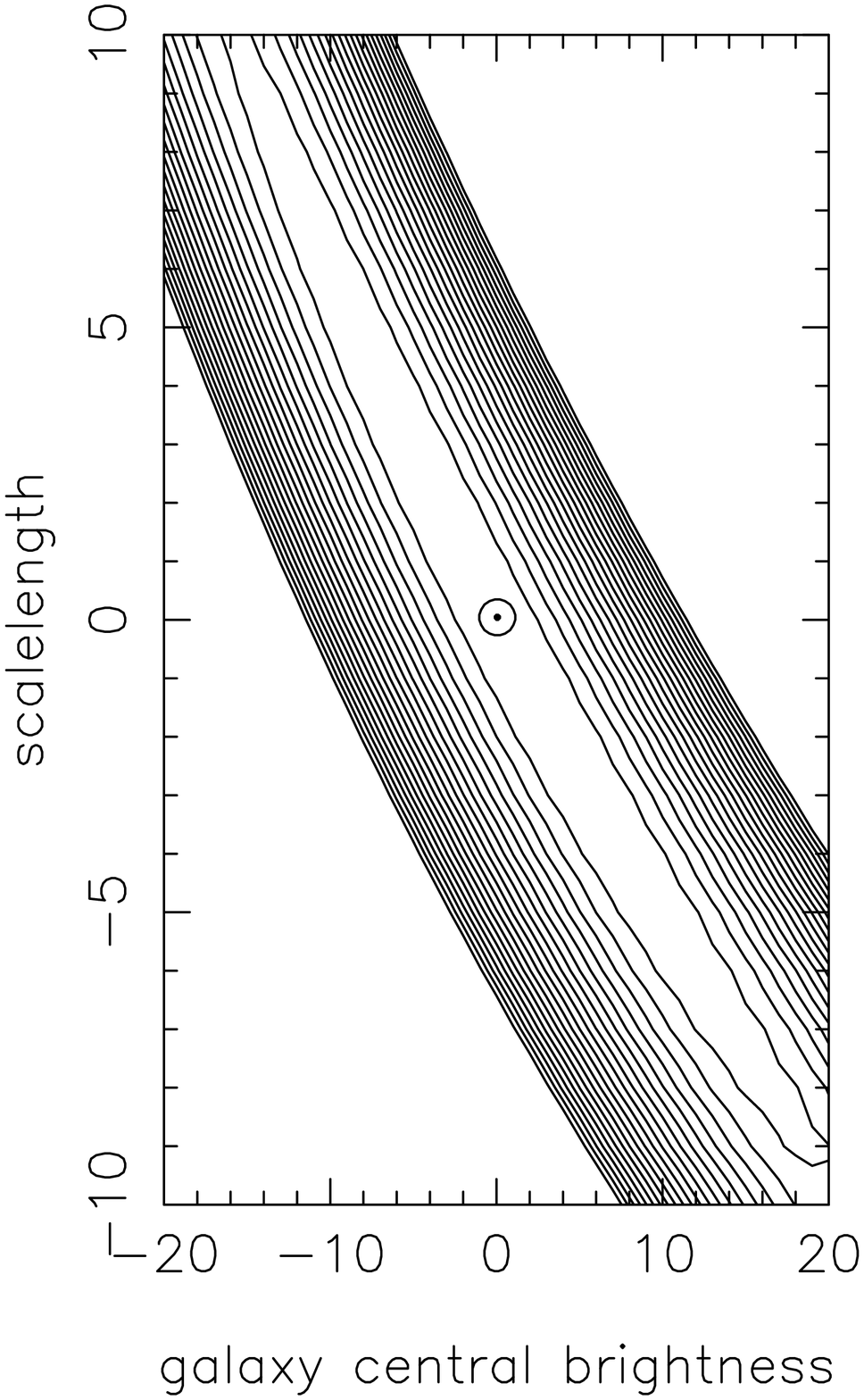}}
\caption{ $\chi^{2}$ contour maps for the radio galaxy 3C234.0 showing 
2-parameter slices
through the 5-parameter hypersurface. Contour levels are spaced at
intervals of $\Delta\chi^{2}=100$ from the minimum $\chi^{2}$ (marked
with $\odot$).}  
\label{conts}
\end{figure}
The main advantage of the two-dimensional modelling technique being
adopted here is that it allows each individual pixel
in the image frame to be included in the model fitting as a degree of
freedom. However, due to the high angular resolution of our data, a
typical model fit to a radius of $12\asec$ will include some 45000
WFPC2 pixels ($\sim5700$ IRCAM3 pixels) in each
$\chi^{2}$ evaluation.  The adopted solution was to utilise
the downhill simplex method (Press et al. 1989) to locate the global
minimum.  Although significantly slower than the conjugate gradient
method used in Taylor et al. (1996), downhill simplex minimization
was chosen for its robustness in finding the true global minimum, and its
insensitivity to the selection of initial conditions. 

To determine the prevalence of
false minima in the hyper-surface, a high-resolution
grid search was undertaken for two objects, with the extent of the
search grid being $\pm20\%$ from the parameter values found by the
minimization routine. The results of the grid search for the radio 
galaxy 3C234.0, which contains a $30\%$ nuclear component, are shown in Fig
\ref{conts} with the position of the global minimum marked with
$\odot$. Two features of this diagram are worthy of comment. Firstly
it can be seen that in both parameter slices the minimum
$\chi^{2}$ value lies at the bottom of a narrow `valley', suggesting
that false minima are not ubiquitous. Secondly, despite the expected 
correlation between galaxy scalelength and
central surface-brightness (Abraham et al. 1992),
the minimization routine has still successfully located the global minimum.

\section{Testing the Modelling Code}
\label{tests}
For the purposes of the full {\sc hst} host galaxy study it was essential to
determine the typical accuracy with which host-galaxy parameters could
be recovered. Due to the nature of host
galaxy work, a data-set for which the host galaxy parameters have been
accurately determine previously does not exist, and consequently, an extensive
programme of tests was undertaken using synthetic data.

To allow a realistic representation of the range of redshifts
displayed by the WFPC2 data the synthetic AGN were constructed at
three redshifts, z=0.1, z=0.2 \,\&\, z=0.3. At each redshift 28
elliptical and 28 disc host galaxies models were constructed. Each group of 28
host galaxies consisted of a range of four possible half-light radii
(5,\,10,\,15 and 20 kpc for $\Omega_{0}=1.0, H_{0}=50$), all of which had
the same integrated luminosity. To each of these four different galaxy
scalelengths a range of seven possible nuclear
components were added, giving $L_{nuc}/L_{host}$ ratios in the range 
$0\rightarrow16$. The
synthetic AGN with no nuclear component were included to check
that the modelling code did not have a bias towards preferring a
nuclear component where none existed.  The final ensemble of models
totalled 168 and covered the full range of parameters likely to occur
in reality.

\subsection{Synthetic Quasar Construction}
\begin{figure}
\setlength{\epsfxsize}{0.4\textwidth}
\hspace{0.4in}
\centerline{\epsfbox{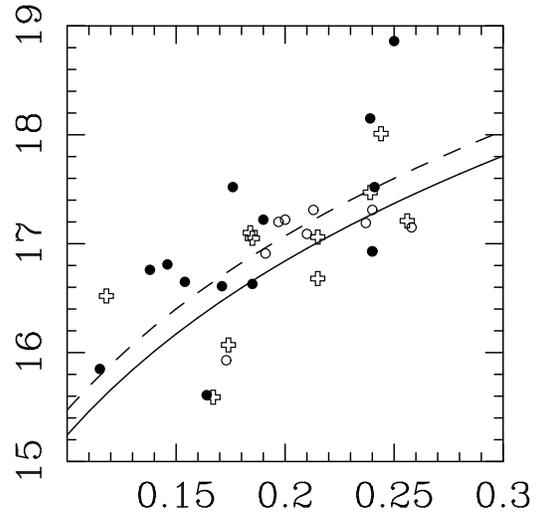}}
\caption{The apparent magnitude versus redshift distribution of the
best-fitting host galaxies of the {\sc hst} sample (Dunlop et al. 2000). 
Shown in the diagram are radio galaxies (crosses), radio-loud quasars (open
circles) and radio-quiet quasars (filled circles). Also
shown is the apparent magnitude of the synthetic host galaxies used for
testing the two-dimensional modelling code (solid line). The dashed
line shows the apparent magnitudes of the synthetic hosts dimmed by
0.27 magnitudes (see text).}
\label{synthgals}
\end{figure}

When constructing the suite of synthetic quasars several steps were 
taken to ensure that these simulations reflected 
the characteristics of actual WFPC2 observations as closely
as possible. The first step was to construct the final synthetic quasar
images from a stack of separate frames in an identical fashion to
the real data. As described in McLure et al. (1999), the actual quasar
observations consist of three deep 600-second exposures which were
complimented by three shorter snap-shot exposures of 5, 26 and 40
seconds, designed to ensure a unsaturated measure of the nuclear
component. The final reduced quasar images consist of
a stack of the three long exposures, with the central regions replaced
by a scaled snap-shot exposure to recover the quasar's full dynamic
range. The construction of the synthetic quasars proceeded in exactly
the same fashion, with individual simulations of each of the six
separate exposures. An appropriate level of background counts were
then added to each of the simulated exposures before they were
processed by the {\sc iraf} routine {\sc mknoise} which simulated the
effects of both shot and read-out noise.

The fudicial luminosity of the simulated host galaxies was chosen to
match the absolute magnitude of the best-fit to {\sc pks} 
2247+14 (M$_{R}=-23.8$), which was typical of the 12 hosts which had 
been observed at the time.  The calculation of the absolute
luminosity of the hosts at the three redshifts was then determined
assuming a typical spectral index of $\alpha=1.5$ for the F675W filter
($f_{\nu}\,\alpha\,\nu^{-\alpha}$), and a cosmology with
$H_{0}=50$kms$^{-1}$Mpc$^{-1}$ and $\Omega_{0}=1$.  The cosmological 
dimming of the AGN point
source assumed a spectral index of $\alpha=0.2$ (Neugebauer et
al. 1987). Figure \ref{synthgals} shows the apparent magnitudes for all 33
host galaxies from the {\sc hst} programme (see Dunlop et al. 2000) 
plotted against redshift. Also shown are the apparent magnitudes of
the synthetic host galaxies. There is a suggestion from Fig
\ref{synthgals} that the
synthetic galaxies are fractionally too bright compared to the data,
making it arguably slightly too easy for the code to recover the 
galaxy parameters from the simulated images. This
is confirmed by the results for the full sample which show the mean
absolute luminosity of the 33 host galaxies to be M$_{R}=-23.53$, 0.27
magnitudes fainter than the synthetic hosts used in the testing
programme. To illustrate this point the dashed line in Fig
\ref{synthgals} shows the apparent magnitude of the synthetic
host galaxies dimmed by a further 0.27 magnitudes, which clearly
provides a better match to the data. However, the overestimate of the 
typical host luminosity should be more that off-set by the much larger 
range of $L_{nuc}/L_{host}$  tackled by the modelling code during
testing. While the average $L_{nuc}/L_{host}$ of the quasars from the
{\sc hst} programme is only $2.6$, the code has been tested with values of 
$L_{nuc}/L_{host}$ in the range $0\rightarrow16$. 

One final measure was taken to improve the realism of the synthetic
quasar images. The empirical PSF which was used during the production
of the synthetic data had an artificial centroiding shift of
$\simeq0.01\asec$ applied to it, significantly greater than the
estimated centroiding error. This precaution was taken in light of the
fact that using the same PSF to convolve both the synthetic images and
the models used in the fitting process is obviously an idealised
situation. 

The results from the modelling of the synthetic quasars are listed in
Tables \ref{ellips}, \ref{discs} and \ref{moreres}. Two features of these
results are worthy of individual comment. Firstly, it can be seen that
for both host morphologies the errors associated with the determination
of all the parameters steadily increase with redshift, as is expected
due to inevitable drop in signal-to-noise. Secondly, with regards to
the host scalelength it can be seen to be significantly easier to
accurately determine this parameter for disc hosts than for
ellipticals. This is again
as is expected considering the different behaviour of the Freeman and
de Vaucouleurs surface-brightness laws in the central
$\simeq1\asec$. The cusp-like nature of the $r^{1/4}$ law at small
radii inevitably leads to greater difficulty in
de-coupling the relative contributions of the host and nuclear
components. 

\noindent
The results of the synthetic data testing can be summarized as follows:
\begin{itemize}
\item{$100\%$ success in host morphology discrimination.}
\item{host flux determination $\geq95\%$ accurate in all cases.}
\item{error in $r_{e}\leq15\%$ out to $z=0.3$ for ellipticals.}
\item{error in $r_{0}\leq3\%$ out to $z=0.3$ for discs.}
\end{itemize}

\noindent
where successful morphological determination refers to a $\Delta
\chi^{2}\ge 25.7$ between the best-fitting model of the correct
morphology, and the best-fitting alternative model, a difference
equivalent to the $99.99\%$ confidence level for a 5-parameter 
fit (Press et al. 1989). Although the high degree of accuracy achieved
in these simulations is impressive, it should be noted that these error
estimates are only valid for the high resolution data provided by {\sc
hst}. As will be seen in Section \ref{tip-tilt}, the errors associated
with ground-based data can be significantly larger.
\begin{table}
\begin{tabular}{cclccccc}
\hline
z&$r_{1/2}$&Nuc&$r_{1/2}$&PA&b/a&Host&Nuc/Host\\
 &/kpc     &Flux   &         &  &   &Flux&Ratio   \\    
\hline
$0.1$&$5$ &2.50&2.40&0.04&0.69&0.37&2.62\\
$0.1$&$10$&1.77&2.90&0.13&0.67&0.83&1.15\\
$0.1$&$15$&1.48&3.09&0.13&0.97&1.26&0.35\\
$0.1$&$20$&1.33&3.44&0.30&1.41&1.86&0.82\\
$0.2$&$5$ &2.15&5.17&0.71&1.33&0.90&2.67\\
$0.2$&$10$&2.52&6.17&1.02&1.84&1.21&1.47\\
$0.2$&$15$&2.25&6.86&0.59&2.20&2.59&0.80\\
$0.2$&$20$&1.78&7.51&0.46&2.77&3.76&2.03\\
$0.3$&$5$ &2.48&8.91&1.51&1.63&1.83&1.73\\
$0.3$&$10$&3.10&9.59&1.63&2.96&1.54&1.43\\
$0.3$&$15$&2.42&10.3&1.13&2.54&3.73&1.48\\
$0.3$&$20$&1.18&9.43&2.31&3.31&4.84&2.29\\
\hline
\end{tabular}
\caption{Results of the two-dimensional modelling tests using
synthetic quasars with elliptical host galaxies. Column 1 gives the
redshift of the quasar. Column 2 gives the actual scalelength of the
simulated host in kpc. Columns 3-8 give the mean percentage error in
the reclaimed value of the relevant parameter. Each value is the mean
for the seven synthetic quasars produced at a particular redshift,
with a particular scalelength (for each redshift and scalelength
combination there where seven different synthetic quasars produced
with $L_{nuc}/L_{host}$=0, 0.5, 1.0, 2.0, 4.0, 8.0 \&\ 16).} 
\label{ellips}
\end{table}
\begin{table}
\begin{center}
\begin{tabular}{cclccccc}
\hline
z&$r_{1/2}$&Nuc&$r_{1/2}$&PA&b/a&Host&Nuc/Host\\
 &/kpc     &Flux   &         &  &   &Flux&Ratio   \\
\hline  
$0.1$&$5$ &1.10&0.53&0.11&0.23&0.16&1.30\\
$0.1$&$10$&0.27&0.44&0.30&0.89&0.73&0.48\\
$0.1$&$15$&0.13&0.43&0.34&1.64&1.10&1.22\\
$0.1$&$20$&0.13&0.70&0.66&2.51&1.43&1.40\\
$0.2$&$5$ &1.60&1.07&0.43&0.80&0.57&1.83\\
$0.2$&$10$&1.32&1.36&0.69&1.97&1.07&1.13\\
$0.2$&$15$&0.48&1.23&0.67&3.16&2.01&2.12\\
$0.2$&$20$&0.48&1.67&0.65&5.39&2.79&2.60\\
$0.3$&$5$ &1.53&1.37&0.67&0.56&0.74&1.73\\
$0.3$&$10$&1.90&1.97&1.47&2.77&1.74&0.75\\
$0.3$&$15$&0.42&1.87&1.20&4.27&2.17&2.10\\
$0.3$&$20$&0.33&1.76&2.11&4.96&2.50&2.67\\
\hline
\end{tabular}
\end{center}
\caption{Results of the two-dimensional modelling tests using
synthetic quasars with disc host galaxies. Columns as Table \ref{ellips}.} 
\label{discs}
\end{table}
\begin{table*}
\begin{center}
\begin{tabular}{lcccccc}
\hline
Host&z&$r_{1/2}$&PA&b/a&Host            &Nuc/Host\\
Type& &         &  &   &Flux $(12\asec)$&Ratio\\
\hline
disc &0.1&0.2$\rightarrow$1.3&0.0$\rightarrow$2.3&0.8$\rightarrow$5.4&0.1$\rightarrow$5.0&0.0$\rightarrow$4.5  \\
elliptical&0.1&1.6$\rightarrow$4.8&0.0$\rightarrow$0.6&0.0$\rightarrow$2.3&0.1$\rightarrow$2.3&0.0$\rightarrow$4.0  \\ 
disc &0.2&0.8$\rightarrow$1.9&0.0$\rightarrow$2.3&0.8$\rightarrow$5.4&0.1$\rightarrow$2.5&0.0$\rightarrow$4.5  \\
elliptical &0.2&2.4$\rightarrow$11.5&0.0$\rightarrow$1.7&0.8$\rightarrow$3.1&0.2$\rightarrow$4.3&0.0$\rightarrow$4.5  \\ 
disc &0.3&0.0$\rightarrow$2.6&0.0$\rightarrow$3.7&0.0$\rightarrow$5.4&0.1$\rightarrow$3.0&0.0$\rightarrow$4.0\\
elliptical &0.3&6.0$\rightarrow$13.7&0.0$\rightarrow$6.3&0.8$\rightarrow$4.6&0.2$\rightarrow$5.0&0.0$\rightarrow$6.8\\  
\hline
\end{tabular}
\end{center}
\caption{The range of percentage errors in the reclaimed values of the
host-galaxy parameters from the synthetic quasar modelling
tests. Columns 1 \&\ 2 detail the actual host-galaxy morphology and
redshift of the synthetic quasars. Columns $3\rightarrow 7$ show the
range in percentage error in the reclaimed parameters from the model
fits to the 28 synthetic quasars constructed at each redshift, with
each of the two host morphologies.}
\label{moreres}
\end{table*}

\section{Beta Parameter Modelling}
\label{betamod}

Due to the success of the programme of tests outlined above, it was
 felt that the
level of information present in the {\sc hst} data justified an
 extension of the modelling code to cover more than just fixed
 elliptical and disc
host galaxies. The modelling code as described so far is able to determine
host galaxy morphology only to the extent that it is elliptical-like or
disc-like. Given that the results of the synthetic quasar testing show
that it is relatively easy to discriminate between
idealised elliptical and disc host galaxies, it is interesting to ask
whether it is also possible to quantify just how similar the host galaxy
surface-brightness distributions are to the classical $r^{1/4}$ or
exponential laws. This question has particular relevance because of
recent studies of the cores of inactive elliptical galaxies using both
{\sc hst} (Lauer et al. 1995) and ground-based imaging (D'Onofrio {\it
 et al.} 1994) which suggest that the
surface-brightness profiles of these galaxies can deviate significantly
from an $r^{1/4}$ law.  It is also of some interest to determine
whether or not the code can differentiate between the $r^{1/4}$ law
and a somewhat flatter relation at ($r\ge
r_{e}$) given the substantial evidence that brightest cluster galaxies
(BCG) display so-called `halos' in their surface-brightness
distributions at large radii (Graham et al. 1996, Schombert 1987).

The Freeman exponential law and the de Vaucouleurs $r^{1/4}$ law can
be thought of as special cases of a more general form of
surface-brightness distribution:
\begin{equation}
\mu(r)=\mu_{o}\rm{exp}\left(-(\frac{r}{r_{o}})^{\beta}\right)
\label{beta}
\end{equation}
\noindent
where the form of the radial profile is governed by the extra free
parameter $\beta$ (Sersic 1968). A further batch of testing was
undertaken to determine whether the value of the this $\beta$
parameter could be recovered with useful accuracy.

\subsection{Testing the Beta Modelling Code}
\label{betatests}
\begin{figure}
\setlength{\epsfxsize}{0.4\textwidth}
\centerline{\epsfbox{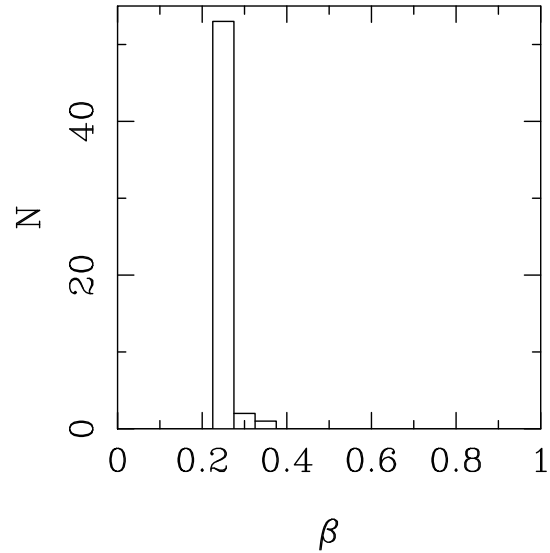}}
\caption{The distribution of beta values recovered by the
$\beta$-modelling code from the synthetic quasars with elliptical host
galaxies at z=0.2 and z=0.3.}
\label{betahist1}
\end{figure}

The question which was asked of the $\beta$-modelling code was
whether, given a sample of synthetic quasars
with idealised de Vaucouleurs host galaxies, could it return a range
of $\beta$ values narrow enough to provide useful extra
information. The accuracy of the $\beta$-modelling code was tested
using the 56 $z=0.2 \,\&\, z=0.3$ synthetic quasars with elliptical hosts from 
Section \ref{tests}. It was not considered necessary to test the 
$\beta$-modelling code
using idealised Freeman disc hosts since the test results presented in
Section \ref{tests} showed that it is substantially more
difficult to recover the parameters of elliptical hosts in all
cases. The success of the
$\beta$-modelling code can be immediately seen from Fig
\ref{betahist1}. Out of the 56 synthetic quasars tested, the $\beta$
value recovered by the modelling code lies in the range
$0.225<\beta<0.275$ in 53 cases.

\section{Combined Disk-Bulge Modelling}
\label{discbulge}

During the latter stages of the analysis of the {\sc hst} host galaxies
presented by Dunlop et al. (2000), it became necessary to extend 
the modelling
code further than has been described above. The $\beta$-modelling of
 four of the radio-quiet quasars from the
the {\sc hst} sample showed that the underlying host galaxy was a hybrid
of both disk and bulge (elliptical) forms. To investigate whether an
improved fit could be achieved with a two-component model the central
model-building algorithm was extended to produce combined disk and
bulge models. During the fitting of these combined models the eight
parameters controlling the form of the galaxy surface-brightness
distributions were left free. In combination with the normalization of
the nuclear component this required the fitting of a total of nine
free parameters. 

With regards to host-galaxy morphology, the clear conclusion from both
 sets of modelling tests
 presented here is that if the host galaxies of the AGN in the {\sc hst} 
imaging study are consistent with standard de Vaucouleurs or
Freeman models, then the two-dimensional modelling code will
successfully discriminate between the two. In addition to this, if the
distribution of $\beta$ values returned from the modelling of the host
galaxies does not show a strong peak around either $\beta=0.25$ or
$\beta=1.0$, then this can be taken as strong evidence the 
host galaxies deviate significantly from the
simple elliptical and disc forms, either because they contain a significant
mix of bulge and disc components, or, for example, because they have been 
distorted by gravitational interaction with a companion.
  
\section{Tip-Tilt Observations of Quasar Host Galaxies}
\label{tip-tilt}
With the arrival of active and adaptive optics systems the prospect of 
near diffraction-limited ground-based
imaging has become a reality. Within the context of the study of
quasar host galaxies this development perhaps has most relevance in
the near-infrared. The inherent advantages of observing host galaxies
in the near-infrared were discussed by Dunlop et al. (1993),
and formed the motivation for the original IRCAM 1 $K$--band imaging of
the {\sc hst} sample (Dunlop et al. 1993; Taylor et al. 1996). The main
limitation of the existing $K$--band
imaging of this sample is the inability to reliably distinguish the 
host galaxy morphology in quasars where the $L_{nuc}/L_{host}\ge5$ due to the
ground-based seeing of $\ge 1\asec$. In
this section the results are presented from a short observing run which
was designed to explore whether the sub-arcsec imaging quality now 
readily attainable with tip-tilt active optics at UKIRT could rectify 
this problem.

\subsection{The Sample}

The observed sample consists of 9 objects (Table \ref{figs}) which
were chosen to fully 
quantify the improvements that could be gained
over the previous $K$--band imaging study. Six of the quasars were taken
from the original 40-object sample described in Taylor et al. (1996), with five
of these also featuring in the new {\sc hst} sample. The remaining three
objects were taken from the sample of Bahcall et al. (1994), and 
were selected because of the original failure to detect a host galaxy from 
$V$--band {\sc hst} imaging (Bahcall it et al. 1994, 1995a). Four of the six 
objects taken from the original 40-object sample (marked with a
$\star$ in Table \ref{figs}) were specifically chosen for this
project because the modelling of the original
IRCAM1 observations was unable to determine the host morphology due
to their high $L_{nuc}/L_{host}$. These objects
presented an ideal opportunity to investigate the advantages to be
gained by confining the nuclear flux to within the central
$0.5\rightarrow 1\asec$ with active optics. The remaining two objects, on
the contrary, were specifically chosen because the model fits from the
previous analysis were regarded as secure, with a strong preference
being shown for one particular host morphology. 
\begin{table}
\begin{center}
\begin{tabular}{lccccl}
\hline
& & & &Time&   \\
Source&Type&$z$&$V$&(mins)&Comment  \\
\hline
0923$+$201$\star$ & RQQ & 0.190 & 15.8 & 72 & Good\\
0953$+$415$\star$ & RQQ & 0.239 & 15.6 & 72 & Good\\
1004$+$130$\star$ & RLQ & 0.240 & 15.2 & 72 & Good\\
1217$+$023        & RLQ & 0.240 & 16.5 & 72 & Good\\
1012$+$008        & RQQ & 0.185 & 15.9 & 72 & EN\\  
1048$-$090$\star$ & RLQ & 0.345 & 17.0 & 72 & PD\\ 
1202$+$281        & RQQ & 0.165 & 15.6 & 36 & Low SNR\\    
1302$-$102        & RLQ & 0.286 & 15.2 & 72 & PD\\    
1307$+$085        & RQQ & 0.155 & 15.1 & 72 & Cloud\\    
\hline
\end{tabular}
\caption{The sample. The first six objects listed are taken from the 
original $K$--band imaging sample (Taylor et al. 1996). The
final three objects have been taken from the sample imaged by Bahcall {\it et
al.} (1994). Column five lists the on-source integration time for each
object. Column six details any problems experienced with the images:
EN = Electronic Noise, PD = Pointing Drift. Redshifts and V magnitudes 
have been taken from Taylor et al. (1996) and Bahcall et al. (1997) 
respectively.}

\label{figs}
\end{center}
\end{table}

\subsection{The Observations}
\label{observations}
The new $K$-band observations were made on 5-7th April 1997 using the
IRCAM 3 infrared camera on the 3.9m United Kingdom Infrared Telescope
({\sc ukirt}) on Mauna Kea, Hawaii. IRCAM 3 is a $256\times256$ InSb
array which was operated in 0.281 arcsec pixel$^{-1}$ mode, providing a
field-of-view of approximately 70$\asec$.  The observing run for this
project was among the first ever to make full use of the tip-tilt
active optics system, producing images with consistent resolution
of $\simeq0.7\asec$. The following observational procedure was used. 

Each object was
observed using a 4-point jitter pattern with each point consisting of
3 minutes of integration broken into 18 co-adds of 10-seconds
duration. This combination was chosen specifically in order to provide
unsaturated but background-limited images. This 12-minute jitter
pattern was repeated six times for
each object, providing a total of 72 minutes of on-source
integration. Considering the
desirability in host galaxy observations of confining the quasar
nuclear flux to as small an angular extent as possible, the quasar
nucleus itself was used as the tip-tilt guide-star in an
attempt to obtain the best possible resolution. In order to provide 
high signal-to-noise measurements of
the IRCAM 3\, PSF, and to calibrate our photometry, observations of
standard stars were taken before and after the completion of each
jitter pattern.
\subsection{Reduction}
\label{reduction}
Following dark-frame subtraction each object was flat-fielded using 
concurrent sky flat-fields produced by a process
of median filtering of the 24 individual 3-minute integrations. Due to
the considerable angular extent of the host galaxies ($\simeq
15\asec$) compared with the IRCAM 3 field-of-view, this median
filtering had to be
performed on a quadrant-by-quadrant basis.  As a result of the jitter
pattern, each quadrant of the array looked at blank sky for 3/4 of the
total integration time for each object.  These frames could be
median filtered without fear of contamination from host galaxy light.
The four flat-field quadrants produced in this fashion were then added
together and normalized to unit median to produce the final
flat-field. After flat-fielding the individual 3-minute frames were 
corrected for the known non-linearity of the IRCAM 3 detector before being
re-registered and stacked to produce the final deep images for
analysis.

\subsection{Image Defects}
\label{defect}

Several problems with the data obtained during this observing run
limited the number of objects which could be successfully analysed
with the modelling code. Firstly, the two highest redshift objects 
(1048-090 \&\ 1302-102)
proved to have nuclear luminosities which were insufficient to provide
stable pointing with the tip-tilt system. The consequent wandering of
the telescope pointing during the observations of these two objects
resulted in final mosaiced images from which it was impossible to
accurately separate the host and nuclear light. Secondly, all of the
images obtained were affected to some extent by bands
of spurious electronic noise, with 1012+008 being the source worst
effected. It proved impossible
to construct a reliable flat-field for this object, 
and it is was therefore dropped from the modelling process.

The data for two more of the nine objects listed in Table \ref{figs} were of
insufficient quality to successfully model the underlying host
galaxies. In the case of 1202+281 this was simply due to the
integration time acquired (36 minutes) providing low signal-to-noise,
while 1307+085 was rejected due to flat-fielding problems produced by 
partial cloud-cover.

The result of these technical difficulties was that only four objects 
yielded data of the high quality required by the
two-dimensional modelling code. However, the remaining four objects 
still allowed the main objectives of the run to be achieved. 
Three of the objects were 
allocated unreliable fits from the previous
$K$--band imaging, and therefore presented a good test of the
improvements to be
gained from the increased resolution of tip-tilt imaging. In addition to
this, the final object (1217+023), was considered to have a reliable
model fit and therefore gave an opportunity to see if the modelling of
the new data was consistent with the previously obtained results.

\subsection{The IRCAM3 Point Spread Function}

As was mentioned
in Section \ref{observations}, standard stars were observed throughout
the observations of each object to provide accurate, concurrent
measures of the PSF.  However, considering that the final deep quasar images
are typically stacks of 24 individual 3-minute frames, in
some cases taken over two nights, there is no guarantee that any
one of the PSF observations will be a good match to the final mosaic.
In an effort to overcome this problem, further tip-tilt PSF
observations were included from a more recent {\sc ukirt} observing
run (Percival et al. 2000), producing a library of 65 high
signal-to-noise PSF observations.  The following procedure was adopted
for choosing the best PSF match for each quasar.

The central nine pixels ($\simeq0.5\asec$) of the quasar image was
compared with each PSF via the $\chi^{2}$ statistic.  The five
individual PSFs which provided the best match to the quasar core were 
then stacked to produce a composite PSF.  The minimization routine 
described in Section \ref{minimum} was then used to find the optimal 
weighting of the individual PSFs to produced the minimum $\chi^{2}$
between the quasar core and the composite PSF.
\subsection{Results}
\label{irc3results}
\begin{table}
\begin{tabular}{ccccccc}
\hline
Source &$\Delta\chi^{2}$&$r_{e}$ /
kpc&K$_{host}$&K$_{nuc}$&b/a&$L_{n}/L_{h}$ \\
\hline
0923+201&27&8.6&13.98&12.48&0.87&3.98\\ 
0953+415&14&8.4&15.18&12.72&0.73&9.62\\ 
1004+130&276&8.5&13.92&12.90&0.85&2.55\\ 
1217+023&187&7.7&14.13&13.62&0.84&1.59\\ 
\hline
\end{tabular}
\caption{Results of the two-dimensional modelling of the IRCAM 3 data. 
Column 2 lists the difference in $\chi^{2}$ between the
best elliptical and disc host galaxy fits (all sources were better
described by an $r^{1/4}$ fit). Column 3 details the effective
radius of the best-fit host galaxy. Columns
4\,\&\,5 list the integrated apparent magnitudes of the host and
nuclear component respectively. Column 6 gives the axial ratio ($b/a$)
for the best-fit host. Column 7 converts the figures of columns
4\,\&\,5 into a nuclear:host ratio. }
\label{irc3res}
\end{table}
\begin{table}
\begin{tabular}{ccccccc}
\hline
Source &$\Delta\chi^{2}$&$r_{e}$ /
kpc&R$_{host}$&R$_{nuc}$&b/a&$L_{n}/L_{h}$ \\
\hline
0923+201&1733&8.2&17.22&15.66&0.98&4.23\\ 
0953+415&91&7.1&18.15&15.19&0.86&15.39\\ 
1004+130&501&8.2&16.93&15.02&0.94&5.78\\ 
1217+023&2359 &9.9&17.31&16.32&0.80&2.49\\ 
\hline
\end{tabular}
\caption{Results of the two-dimensional modelling of the {\sc hst}
data. Columns are as Table \ref{irc3res} with the exception of columns
4\,\&\,5 which here list the $R$-band magnitudes of the host and
nuclear component respectively.}
\label{hstresults}
\end{table}

\begin{figure*}
\centerline{\epsfig{file=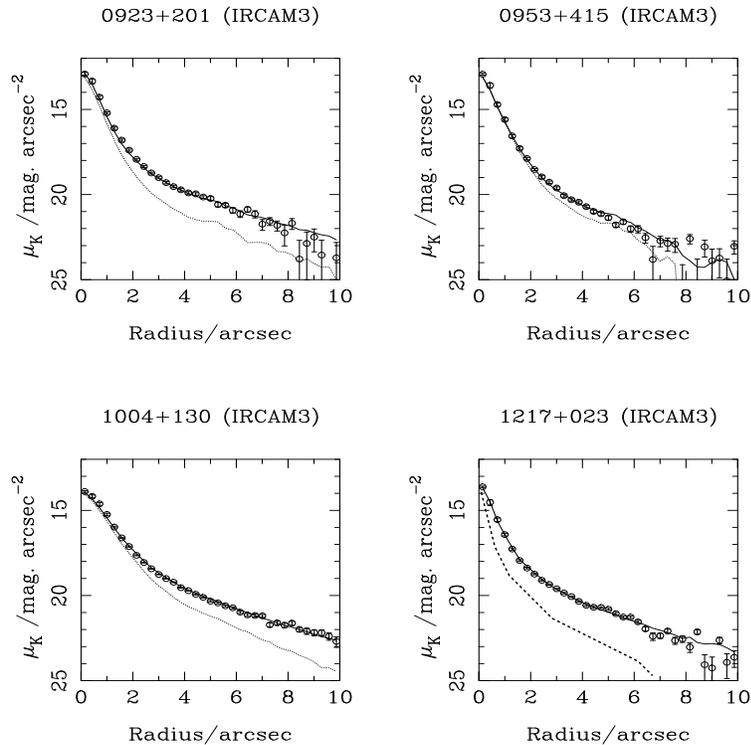,width=10.0cm,angle=270,clip=}}
\caption{The surface-brightness profiles resulting from the
two-dimensional modelling of the IRCAM3 data. Shown in the
figure are the data (open circles), the best-fit model (solid line)
and the best-fit nuclear component (dashed line).}
\label{prof}
\end{figure*}

The results from the two-dimensional modelling of the four quasars
which did not suffer from image degradation are listed in Table
\ref{irc3res}. Surface-brightness profiles extracted from the
two-dimensional model fits are shown in Fig \ref{prof}. Also
shown in Table \ref{hstresults} are the results from the modelling of
the {\sc hst} $R$--band imaging of the four objects. These results are identical
to those presented in McLure et al. (1999) and Dunlop {\it et
al.} (2000), and are repeated here simply for ease of comparison with the new 
$K$-band results.

\subsection{Host Morphologies}

All four of the quasars are found to lie in elliptical host galaxies,
 just as they were from the {\sc hst} imaging, despite 0923+201 and 0953+415 
being radio-quiet. The morphological
decision for 0953+415 is the least clear-cut, formally only $2\sigma$,
 as expected
considering that it has by far the largest value of
 $L_{nuc}/L_{host}$. As well as being in agreement
 with the {\sc hst} modelling results, the morphology decisions for the new $K$-band
 images of 0953+415 and 0923+201 are fully consistent with the predictions of
Taylor et al. (1996) that luminous RQQs are likely to have early-type host
 galaxies, with 0953+415 and 0923+201 being the most optically
 luminous RQQs in the full {\sc hst} sample.

\subsection{Scalelengths}
\begin{figure}
\centerline{\epsfig{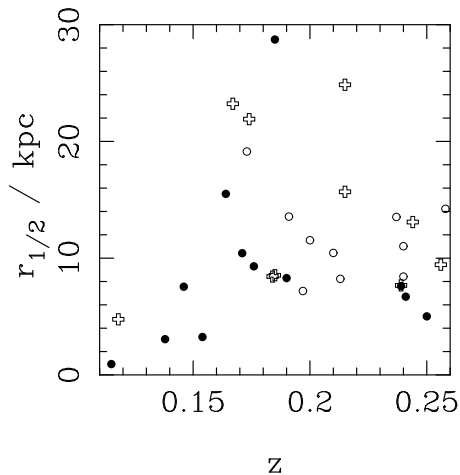}}
\caption{The $r_{1/2}-z$ distribution of the best-fitting host models
of the RG (crosses), RQQ (filled circles) and RLQ (open circles) sub-samples.}
\label{re_z}
\end{figure}
An investigation of Tables \ref{irc3res} and \ref{hstresults} shows
that the best-fit scalelengths from the two independent sets of data
are in excellent agreement. In both wavebands the hosts are found to
have large, and very similar scalelengths, spanning a range of
less than 3 kpc.

The general result that these quasars are hosted by large
galaxies with scalelengths of $r_{e}\simeq10$ kpc is consistent with
the conclusions of Taylor et al. (1996). However, with the exception of
0953+415, the scalelength results presented here are not in good
agreement on an individual object-by-object basis, differing by more than a
factor of two in all cases. The reason for this most likely lies in
the fact that, again with the exception of 0953+415, the best-fitting
$L_{nuc}/L_{host}$ from the modelling of the IRCAM 1 data are
$2\rightarrow4$ times greater than those found using IRCAM 3. Given that
an overestimate of the unresolved nuclear
component will artificially lower the central surface-brightness of
the best-fitting host-galaxy model, the requirement to match the
host-galaxy flux at large radii will therefore lead to an inevitable 
overestimate of
the host-galaxy scalelength. This conclusion is strengthened by a
comparison of the {\sc hst} scalelength results with the
corresponding figures from Taylor et al.  for all 33-objects in the
{\sc hst} sample (Dunlop et al. 2000). In the majority of cases there
is a consistent
bias in the previous IRCAM 1 $K$--band modelling towards finding 
larger galaxies
with a lower central surface-brightness, although the difference is
generally not as great as that found for the four quasars presented here,
which are amongst the most luminous in the sample. The detection of
this bias confirms that scalelength values derived from data without 
sub-arcsec resolution should be regarded with some caution.

If viewed in isolation it could appear of some concern that the
best-fit scalelengths of all four
host galaxies should be so similar, given that three of them
have identical redshifts. However, the possibility that there is some
bias in the modelling can be firmly rejected for a number of
reasons. Firstly, during the testing
of the modelling code (Section \ref{tests}) there was no such bias
detected,
 with the code successfully discriminating between scalelengths in the range
$5\rightarrow20$ kpc. Secondly, the close agreement
between the optical and infrared scalelengths derived for these four 
objects provides further reassurance that there is no correlation 
between scalelength and redshift, given that Fig
\ref{re_z} shows that no such correlation exists in the scalelengths
determined for the full {\sc hst} sample (Spearmann rank test probability of
$p=0.52$).
\subsection{Host Luminosities and Colours}

\begin{table}
\begin{center}
\begin{tabular}{cccccc}
\hline
Source& M$_{K}$ & $L/L^{\star}_{K}$ &M$_{R}$ & $L/L^{\star}_{R}$&$R-K$ \\
\hline
0923+201&$-$26.21&4.3&$-$23.25&2.4&3.0\\

0953+415&$-$25.48&2.2&$-$22.82&1.6&2.7\\

1004+130&$-$26.75&7.1&$-$24.10&5.2&2.7\\

1217+023&$-$26.54&5.9&$-$23.71&3.7&2.8\\
\hline
\end{tabular}
\end{center}
\caption{Luminosities and colours for the four best-fitting host
galaxies. Column 2
lists the absolute integrated $K$--band magnitudes. Column 3 restates
the absolute magnitudes in term of $L^{\star}$. Columns 4 \&\ 5 give
the equivalent figures from the R--band {\sc hst} modelling. Column 6 gives
the resulting $R-K$ colour for the host galaxies.} 
\label{kmags}
\end{table}

The integrated absolute $K$--band and $R$--band magnitudes of the
 best-fitting host galaxy models
are presented in Table \ref{kmags}. The absolute magnitudes have
been calculated using $k$-corrections assuming a spectral index
$\alpha=0.0$ and $\alpha=1.5$ for the $K$- and $R$--band respectively
 ($f_{\nu}\,\alpha\,\nu^{-\alpha}$). The calculation of
the host luminosities in terms of $L^{\star}$ have been performed
taking $M^{\star}_{K}=-24.6$ (Gardner et al. 1997) and 
$M^{\star}_{R}=-22.3$ (calculated by converting the value of 
$M^{\star}_{R}=-21.8$ (Lin et al. 1996) to an integrated
 magnitude). Again it can be seen that these results confirm
the findings of the {\sc hst} $R$--band imaging, and the previous $K$--band
imaging (Taylor et al. 1996), that the quasar hosts are all luminous
galaxies with  $L\geq2L^{\star}$. The fact that in this small group of
objects the RLQs are substantially more luminous than the
RQQs should not be taken as significant since, 1004+130 and 1217+023 are
found to be brighter than average for RLQs from the $R$--band modelling, while
0953+415 has the largest $L_{nuc}/L_{host}$ ratio 
in the entire 33-object sample.  

The excellent agreement of the model results from the $K$- and
$R$--band imaging in terms of host morphology and scalelength allows
an accurate measurement of the host galaxy optical-infrared colour to
be made. The integrated, rest-frame, $R-K$ colours listed in Table
\ref{kmags} can be seen to be perfectly consistent with that expected
from an old passively evolving stellar population formed at
$z\ge3$. This conclusion is confirmed by the $R-K$ colours of the full
33-object sample, where comparison with spectral synthesis models
indicates that none of the AGN host galaxies have stellar populations
younger that $\sim7$ Gyr at z=0.2 (Dunlop et al. 2000). A further
conclusion which can be drawn from the red colours of these host
galaxies is that any star formation associated with the AGN activity
must be confined to the central few kpc, and not widely distributed
throughout the body of the host galaxy.
\section{Conclusions}

The results of applying our two-dimensional modelling technique 
to the full range of simulated quasar images provides confidence 
that, at least at $z \simeq 0.1 - 0.3$ all derived host galaxy parameters
can be trusted to a typical accuracy of better than 10\%. This is perhaps not 
surprising for parameters such as host luminosity, position angle, and axial
ratio, but accurate scalelength determination has proved particularly
difficult in virtually all previous 
studies of quasar hosts. The reliability of our derived scalelengths is 
further re-inforced by the virtually identical values yielded by the 
analysis of the infrared tip-tilt images discussed in the previous 
section. Indeed the fact that such similar scalelengths have resulted from
analysis of images taken at different wavelengths, with different 
plate-scales, and using very different telescopes with very different PSF
complications is undeniably impressive.
It is this accuracy of scalelength determination that has enabled us, for the 
first time, to demonstrate that the hosts of RQQs, RLQs and RGs follow a 
Kormendy relation with slope and normalization identical to that displayed
by `normal' low-redshift massive ellipticals (McLure et al. 1999; Dunlop et al. 2000).
 
The analysis of the new, higher-resolution (cf Dunlop e al. 1993) infrared
images of the 4 quasars imaged with IRCAM3 indicates that galaxy scalelength
is (perhaps unsurprisingly) essentially unchanged between the $R$ and $K$
wavebands, and that the {\sc hst}-derived scalelengths should be trusted in 
preference to those derived by Taylor et al. (1996) from the IRCAM1 images
of Dunlop et al. (1993). In at least some cases the original infrared-based 
host-galaxy scalelengths appear to have been over-estimated from the 
IRCAM1 images, although not sufficiently to alter the 
basic conclusions of Taylor et al. (1996). 
This has been taken into account in a re-analysis of the IRCAM1
data undertaken by Dunlop et al. (2000) in order to derive reliable $R-K$
colours for {\it all} the host galaxies in the {\sc hst} sample. 

\section{Acknowledgements}
We thank the referee for comments which improved a number of aspects
 of this paper. The United Kingdom Infrared Telescope is operated by 
the Joint Astronomy Centre on behalf of the U.K. Particle Physics and 
Astronomy Research 
Council. Based on observations with the NASA/ESA Hubble Space Telescope, 
obtained
at the Space Telescope Science Institute, which is operated by the
Association of Universities for Research in Astronomy, Inc. under NASA
contract No. NAS5-26555.
This research has made use of the NASA/IPAC Extragalactic Database (NED)
which is operated by the Jet Propulsion Laboratory, California Institute
of Technology, under contract with the National Aeronautics and Space
Administration.
MJK acknowledges the award of a PPARC PDRA, and also acknowledges
support for this work provided by NASA through grant numbers O0548 
and O0573 from the Space Telescope Science Institute, which
is operated by AURA, Inc., under NASA contract NAS5-26555. RJM 
acknowledges a PPARC studentship.

\section{References}
\noindent
Abraham R.G., Crawford C.S., McHardy I.M., 1992, ApJ, 401, 474\\
Bahcall J.N., Kirhakos S., Schneider D.P., 1994, ApJ, 435, L11\\
Bahcall J.N., Kirhakos S., Schneider D.P., 1995a, ApJ, 447, L1\\
Biretta J.A., et al., 1996, WFPC2 Instrument Handbook Version 4.0\\
Boyce P.J., et al., 1998, MNRAS, 298, 121\\
De Vaucouleurs G., Capaccioli M., 1978, ApJS, 40, 699\\
D'Onofrio M., Capaccioli M, Caon N., 1994, MNRAS, 271, 523\\
Dunlop J.S., Taylor G.L., Hughes D.H., Robson E.I., 1993, MNRAS, 264, 455\\
Dunlop J.S., McLure R.J., Kukula M.J., Baum S.A., O'Dea C.P., 2000, MNRAS, 
submitted\\
Freeman K.C, 1970, ApJ, 160, 811\\
Gardner J.P., Sharples R.M., Frenk C.S., Carrasco B.E., 1997, A\&AS, 190, 
4303\\
Graham A., Lauer T.R., Colless M., Postman M., 1996, ApJ, 465, 534\\
Hooper E.J., Impey C.D., Foltz C.B., 1997, ApJ, 480, L95\\
Hutchings J.B., Morris S.C., Gower A.C., Lester M.L., 1994, ApJ, 429, L1\\
Krist J., 1998, TinyTim User Manual\\
Lauer T.R., et al., 1995, AJ, 110, 2622 \\
Lin H., Krishner R.P., Shectman S.A., Landy S.D., Oemler A., Tucker D.L., Schechter P.L., 1996, ApJ, 464, 60\\
McLure R.J., Kukula M.J., Dunlop J.S., Baum S.A., O'Dea C.P., Hughes D.H., 
1999, MNRAS, 308, 377\\
Neugebauer G., Green G.F., Matthews K., McGill J., Scoville N., Soifer B.T.,
 1987, ApJS, 93, 1057\\
Percival W., Miller L., McLure R.J., Dunlop J.S.,  2000, MNRAS, 
astro-ph/0002199\\
Press W.H., 1989, ``Numerical Recipes'', Cambridge University Press\\
Schombert J.M., 1987, ApJS, 64, 643\\
Sersic J.L., 1968, Atlas de Galaxies australes. Observatorio
Astronomico Cordoba\\
Taylor G.T., Dunlop J.S., Hughes D.H., Robson E.I., 1996, MNRAS, 283, 930\\
\end{document}